%% file: paper.tex
\documentclass[sigconf]{acmart}

\usepackage{graphicx}
\usepackage{todonotes}
\usepackage{soul}

\newcommand{\nameCyberRange}{GothX}

\usepackage{soul}

\AtBeginDocument{%
  }

\copyrightyear{2024}
\acmYear{2024}
\setcopyright{rightsretained}
\acmConference[CSET 2024]{Workshop on Cyber Security Experimentation and Test}{August 13, 2024}{Philadelphia, PA, USA}
\acmBooktitle{Workshop on Cyber Security Experimentation and Test (CSET 2024), August 13, 2024, Philadelphia, PA, USA}
\acmPrice{}
\acmDOI{10.1145/3675741.3675753}
\acmISBN{979-8-4007-0957-9/24/08}

\begin{document}

\title{GothX: a generator of customizable, legitimate and malicious IoT network traffic}
\author{Manuel Poisson}
\email{manuel.poisson@irisa.fr}
\orcid{0009-0004-1479-4953}
\affiliation{%
  \institution{Amossys/CentraleSupélec/CNRS/Univ. Rennes/IRISA}
  \city{Rennes}
  \country{France}
}
\author{Rodrigo Matos Carnier}
\email{rodrigo_carnier@nii.ac.jp}
\orcid{0000-0002-0843-3033}
\affiliation{%
  \institution{NII}
  \city{Tokyo}
  \country{Japan}
}
\author{Kensuke Fukuda}
\email{kensuke@nii.ac.jp}
\orcid{0000-0001-8372-2807}
\affiliation{%
  \institution{NII/Sokendai}
  \city{Tokyo}
  \country{Japan}
}

\renewcommand{\shortauthors}{Poisson M. et al.}

\begin{abstract}

\input{abstract}
\end{abstract}



\maketitle

\section{Introduction}

\input{intro}

\section{Related works} \label{sec:related_works}

\input{related_works}

\section{Architecture design for customization and extra improvements} \label{sec:archi}

\input{general_archi}

\section{Use cases} \label{sec:use-case}

\subsection{MQTTset replication}\label{subsec:mqttset}

\input{mqttset}

\input{attack_scenario}

\section{Discussion} \label{sec:eval}

\input{scalable_repro}

\section{Conclusion} \label{sec:conclu}

\input{conclu}

\begin{acks}
Manuel Poisson is supported by the NII internship program, the Collège doctoral de Bretagne, and the research team \href{https://team.inria.fr/pirat/}{PIRAT$\setminus$');}
This work is partly supported by JST CREST JPMJCR21M3.
\end{acks}

\bibliographystyle{ACM-Reference-Format}
\bibliography{mybib}

\end{document}

%% file: abstract.tex
In recent years, machine learning-based anomaly detection (AD) has become an important measure against security threats from Internet of Things (IoT) networks. Machine learning (ML) models for network traffic AD require datasets to be trained, evaluated and compared. Due to the necessity of realistic and up-to-date representation of IoT security threats, new datasets need to be constantly generated to train relevant AD models. Since most traffic generation setups are developed considering only the author’s use, replication of traffic generation becomes an additional challenge to the creation and maintenance of useful datasets. In this work, we propose GothX, a flexible traffic generator to create both legitimate and malicious traffic for IoT datasets. As a fork of Gotham Testbed, GothX is developed with five requirements: 1) easy configuration of network topology, 2) customization of traffic parameters, 3) automatic execution of legitimate and attack scenarios, 4) IoT network heterogeneity (the current iteration supports MQTT, Kafka and SINETStream services), and 5) automatic labeling of generated datasets. GothX is validated by two use cases: a) re-generation and enrichment of traffic from the IoT dataset MQTTset,and b) automatic execution of a new realistic scenario including the exploitation of a CVE specific to the Kafka-MQTT network topology and leading to a DDoS attack. We also contribute with two datasets containing mixed traffic, one made from the enriched MQTTset traffic and another from the attack scenario. We evaluated the scalability of GothX (450 IoT sensors in a single machine), the replication of the use cases and the validity of the generated datasets, confirming the ability of GothX to improve the current state-of-the-art of network traffic generation.

%% file: intro.tex
The deployment of billions of IoT devices has changed the landscape of network security. Their platform heterogeneity creates new surfaces of attack, while their high maintenance cost by sheer volume discourages manufacturers from implementing many security measures~\cite{bagchi2020, asplund2016}. This combination has led to increased security breaches and large-scale DDoS attacks~\cite{kambourakis_mirai_2017}. In this arms race, the increase of malicious activity has also increased their footprint on network traffic, which in turn increased the efficacy of network traffic anomaly detection -- particularly ML-based AD~\cite{eltanbouly2020, kdd99}.

Performances of ML models depend on several factors, but the most important is the quality of the data used in their training. For most studies, datasets are the main source of data~\cite{keersmaeker2023, ring2019}. To effectively detect current threats to IoT networks, datasets must contain a diversity of data from recent and realistic attacks. Moreover, to minimize false alerts, datasets should also include diverse and realistic samples of legitimate traffic. Nevertheless, some public datasets include only legitimate or only malicious traffic~\cite{shirsath_caida_2023}. Generating only legitimate traffic that properly represents IoT heterogeneity is already challenging, but generating malicious IoT traffic that includes a good representation of IoT threats may not even be feasible. An additional issue is the necessity of correct labeling of data to train the ML model. 
However, some mixed datasets are just captured traffic, leaving the labeling to be made by the user himself. To compound all this, continuous deployment of new IoT technologies turns IoT datasets outdated fast. As a consequence, mixed and heterogeneous IoT traffic datasets, properly labeled and still relevant, are constantly in need.

To alleviate these problems, the training of ML-based traffic AD may use more than one dataset, but it may be challenging to merge and balance their data. In the worst case, existing datasets cannot match the typical traffic (either legitimate or malicious) of the target network where the AD solution will be deployed. If so, either live data must be collected -- which poses the problem of availability of relevant malicious traffic, to prepare the model for future attacks -- or a traffic generator must be used to create a dedicated dataset~\cite{anande2023}.

Preparing a testbed from scratch to generate one's own data requires significant time and expertise. Ideally, existing traffic generators should be adapted to IoT traffic and used instead (Section~\ref{sec:related_works}). However, setups used to study traffic and generate public datasets are rarely available themselves. When they are, their implementation may not be flexible enough for the needs of new experiments. Another problem is the generated traffic. The testbed may be public and easy to apply in other studies, but the generated traffic may lack the properties described above. For instance, Gotham~\cite{saez-de-camara_gotham_2023} produces a static dataset mixing legitimate and malicious traffic.
However, the output is hard to modify and Gotham provides no way to automatically assign labels to the traffic. %

To overcome these issues, we provide \nameCyberRange{}, a traffic generator designed to generate customized and labeled datasets of IoT network traffic. It is an improved fork of the replicable IoT testbed for security experiment Gotham. \nameCyberRange{} can automatically execute scenarios in a customized topology (e.g. number of IoT sensors). Simulated scenarios present customizable parameters of operation that generate traffic with more varied features (e.g. volume and frequency of messages customized for each node) (Section~\ref{sec:archi}). We exemplify its capabilities with two use cases  (Section~\ref{sec:use-case}): a) regeneration and enrichment of mixed IoT traffic from the dataset MQTTset, and b) generation of mixed IoT traffic in a custom multi-step attack scenario that results in a DDoS. 
This scenario is automatically executed in a realistic heterogeneous IoT network setup configured with MQTT and Kafka services (a real-world configuration found in the Japanese academic network SINET~\cite{sinet2024, kurimoto2023}).
It consists of the execution of a multi-step attack that compromises a network of sensors and generates a DDoS attack.

Moreover, \nameCyberRange{} is open-source and made publicly available on GitHub\footnote{GothX and the datasets are available at \href{https://github.com/fukuda-lab/GothX}{https://github.com/fukuda-lab/GothX}}. 
 A last contribution of our work is the publication of two new datasets (Section~\ref{sec:use-case}), containing the traffic of the use-cases above. %
 In the process of creating these datasets, we evaluated and validated the scalability and replicability of our solution (Section~\ref{sec:eval}).

In summary, the main contributions of our work are:
\begin{enumerate}
    \item \nameCyberRange{}: a public, easy-to-use and highly customizable IoT traffic generator, capable of producing rich IoT network traffic. 
    \item Critically desired properties of dataset generation, including mixed legitimate and malicious traffic, customizable traffic properties, heterogeneous network setup with MQTT and Kafka services, and automatic labels generation for datasets.
    \item Two labeled datasets: the replication of MQTTset with some improvements and a new dataset with legitimate and malicious traffic based on our attack scenario.
\end{enumerate}

%% file: related_works.tex
Considering the increasing relevance of IoT networks and the compatibility issues that many applications for conventional networks experience when ported to IoT, many simulation tools have been created to facilitate the development of IoT applications. Consequently, surveys on IoT simulation tools are being steadily published~\cite{singh_survey_simulation_2022, patel_simulators_2019, zhu_survey_2022}. They demonstrate the current diversification of IoT tools, which can be roughly divided in simulators, emulators and testbeds according to the trade-off between scalability and realism of simulation. %

However, %
surveys of tools dedicated to traffic generation cannot be found in the literature except for the specific case of GAN traffic generators~\cite{anande2023}. 
This is a strong indicative that there are not many available tools dedicated to this task. 
To better contextualize our contributions to the field, we provide in the section below a short review of existing traffic generators and a description of our design goals with respect to the state-of-the-art.

\subsection{Existing traffic generators}

Traffic generators can be divided into three main categories~\cite{zhu_survey_2022}:  hardware-\allowbreak based, simulation-based, and hybrids. Hardware-\allowbreak based tools are made of real IoT devices. They allow testing interactions between devices in the biggest degree of realism. However, it is difficult and expensive to deploy a real network just for traffic generation, and scalability becomes a prohibitive limitation. 
FIT IoT-LAB~\cite{fleury_fit_2015} tries to mitigate some of these problems by designing rooms with a rich set of IoT devices having different functionalities and providing remote access to researchers willing to perform experiments in these rooms. 
However, any significant modification of the experiment by users is not feasible.
On the other hand, simulation-based tools like CupCarbon-Lab~\cite{bounceur_cupcarbonlab_2018} and Gotham~\cite{saez-de-camara_gotham_2023} rely solely on the memory and computational power of a simulation setup to represent as many devices as possible, with the trade-off of comparatively reduced realism. 
Nevertheless, with recent advances in computation and virtualization, simulation-based tools currently offer many benefits in terms of scalability and a much smaller cost of deployment, with the trade-off of smaller realism. 
However, since the realism of hardware-based tools is high concerning devices but very low with respect to the size and topology of IoT networks, virtualization brings a realistic aspect to the traffic generation with respect to the network topology and deployed services.
Finally, hybrid tools like the IoTEP~\cite{terroso_saenz_hybrid_testbed_2019}, combine both real and virtual devices, trying to balance their advantages and limitations.

Simulation-based traffic generators can also differ regarding the process of packet generation. Leaning toward scalability and simplification of simulation are packet synthesizers, like IoT-Flock~\cite{ghazanfar_iot-flock_2020}. They craft custom packets without a stack of network protocols and simulated devices, directly specifying parameters like headers, payloads and flags. Although it significantly increases the flexibility and volume of generated data, the realism and coherence of crafted data can deteriorate significantly and compromise the utility of the data for training ML-based AD solutions. On the other side of the spectrum are testbeds, which virtualize devices to a big degree, create networks with realistic topologies and traffic controllers like servers and switches, and generate traffic as it would be done by actual hardware. This is the approach chosen for GothX.

Gotham~\cite{saez-de-camara_gotham_2023} is an open-source traffic generator that leverages the popular open-source network emulator GNS3~\cite{gns3_book} to simulate IoT devices and generate datasets in a replicable manner. Virtualization is facilitated and automated with Dockerfiles and virtual machines (VMs), while scripts interact with a GNS3 server using the REST API to execute simulation scenarios. The scenario provided by Gotham generates both legitimate and malicious IoT traffic. Legitimate traffic is created using popular IoT devices like MQTT and CoAP. %
Malicious traffic is created using the automatic launch of attack scenarios from well-known threats, like the botnet Mirai~\cite{kambourakis_mirai_2017}.

To the best of our knowledge, Gotham is the only simulation-based testbed for IoT security that provides open-source and free access to its code. Therefore we chose to develop GothX on top of Gotham functionalities, extending it to produce rich mixed network traffic and automatically label the created dataset.

\subsection{Missing features in the state-of-the-art}

\begin{table}[htpbt]
    \caption{Extended features from Gotham to \nameCyberRange{}}
    \label{table:comparison-gothx-gotham}
    \begin{tabular}{lcc}
        \hline
        \multicolumn{1}{|l|}{\textbf{Features}}                & \multicolumn{1}{c|}{\textbf{Gotham}} & \multicolumn{1}{c|}{\textbf{GothX}} \\
        \hline
        \multicolumn{1}{|l|}{Open-source}                      & \multicolumn{1}{c|}{\checkmark}      & \multicolumn{1}{c|}{\checkmark}     \\ \hline
        \multicolumn{1}{|l|}{Legitimate + malicious traffic}   & \multicolumn{1}{c|}{\checkmark}              & \multicolumn{1}{c|}{\checkmark}               \\ \hline
        \multicolumn{1}{|l|}{Virtualization (Docker + VM)}     & \multicolumn{1}{c|}{\checkmark}              & \multicolumn{1}{c|}{\checkmark}               \\ \hline
        \multicolumn{1}{|l|}{Automatic network initialization} & \multicolumn{1}{c|}{\checkmark}              & \multicolumn{1}{c|}{\checkmark}               \\ \hline
        \multicolumn{1}{|l|}{Replicable results}             & \multicolumn{1}{c|}{\checkmark}      & \multicolumn{1}{c|}{\checkmark}     \\ \hline
        \multicolumn{1}{|l|}{Labeled data}                     & \multicolumn{1}{c|}{}                & \multicolumn{1}{c|}{\checkmark}     \\ \hline
        \multicolumn{1}{|l|}{Customizable node behavior}     & \multicolumn{1}{c|}{}                & \multicolumn{1}{c|}{\checkmark}     \\ \hline
        \multicolumn{1}{|l|}{MQTT service}                     & \multicolumn{1}{c|}{\checkmark}                & \multicolumn{1}{c|}{\checkmark}     \\ \hline
        \multicolumn{1}{|l|}{CoAP service}                     & \multicolumn{1}{c|}{\checkmark}      & \multicolumn{1}{c|}{}               \\ \hline
        \multicolumn{1}{|l|}{MQTT-Kafka service}               & \multicolumn{1}{c|}{}                & \multicolumn{1}{c|}{\checkmark}     \\ \hline
        \multicolumn{1}{|l|}{Accompanying ready-made datasets}               & \multicolumn{1}{c|}{}                & \multicolumn{1}{c|}{\checkmark}     \\ \hline

    \end{tabular}
\end{table}

Gotham made a big contribution to the coordination of efforts in network traffic dataset generation, providing a replicable environment with an automatic setup of network topology and simulation scenario, packaged in an open-source tool that is well documented. To take the next step in the state of the art of traffic generation, we identified three open problems in Gotham: a) The network simulated scenario is not very customizable and has fixed behavior for all IoT sensors; b) it does not generate a labeled dataset, and c) some network traffic typically found in IoT is not represented in Gotham. Below we present a discussion of the impact of these open problems.

In a traffic generator, low customization of network simulation and device behavior translates directly to the quality of the created dataset. By simulating multiple network topologies and patterns of communication, bigger diversity is introduced in the network traffic (a necessity even bigger for intrinsically heterogeneous IoT networks), which is important to train ML-based AD solutions with high precision and recall. 
The ability to easily create different datasets using the same tool is beneficial not only for the creation of such AD models, but also for the study of feature selection itself. 
Explainable artificial intelligence (XAI)~\cite{moustafa_xai_2023} is one example of research field that could leverage the abundance of variable data. Another example is the improvement and update of public datasets, which is difficult due to the lack of availability of tools for dataset replication. \nameCyberRange{} provides an easy customization in its configuration files of a simulation. Since Gotham hard-codes these aspects of simulation in the testbed itself, it is difficult to generate many different scenarios and consequently different traffic data.

Unlabeled datasets are problematic for numerous reasons. Besides requiring significant additional work of manual labeling by users, errors in labeling are much more likely to occur due to the unfamiliarity with the conditions of simulation that created the dataset. 
Moreover, unbalanced data is likely to escape the attention of the dataset creators before it goes public, as is the chance of duplicated packets. 
To provide an internal degree of documentation in the \nameCyberRange{}-generated datasets themselves, we devised a method of labeling the datasets in the same workflow of the simulation scenario. More details will be provided in Section~\ref{sec:archi}.

Regarding the representation of different IoT traffic, we highlight the importance of increasing the heterogeneity of simulated traffic in more than one way. Besides the configuration of network topology and nodes' operation, GothX has additional communication services with their own servers in its simulation scenarios: 
Kafka service, a popular event-streaming platform; 
a Kafka-MQTT connection server, that consolidates traffic data from multiple MQTT networks; 
and SINETStream service~\cite{takefusa_sinetstream_2021}, which is built on top of the services above and is an example of a real-world IoT application.

A final commentary is the absence of different public datasets generated by Gotham.
Since Gotham is configured to produce the same type of traffic, it generates the same dataset when executed. 
We highlight the new features of traffic diversity in GothX by providing two essentially different datasets regarding malicious traffic.
Both are automatically labeled. More details on their differences will be presented in Sections~\ref{subsec:mqttset} and~\ref{subsec:attack}.

Table~\ref{table:comparison-gothx-gotham} summarizes the difference in implemented features between Gotham and \nameCyberRange{}.

%% file: general_archi.tex
This paper presents \nameCyberRange{}, an improved fork of Gotham designed to generate customized labeled datasets.
Like Gotham, we use a real topology where nodes interact with each other automatically.
Agents running on nodes emit legitimate network traffic to other nodes that react accordingly.
On top of that, malicious traffic is produced by an attack agent.
Usage of such real topology in \nameCyberRange{} allows future addition of new nodes interacting with nodes already present.
For example, one could integrate to \nameCyberRange{} a node with an IDS to evaluate the online performances of the IDS.

One of \nameCyberRange{}'s advantages is that it adds interactions using Kafka.
Furthermore, the labeling process of \nameCyberRange{}'s datasets is automated.
However, the main value of \nameCyberRange{} is the easy customization of the generated datasets.
Customization leads to a large number of settings' combinations which favors the diversity of generated datasets.
Thanks to the easy customization of topology and scenario, \nameCyberRange{} can provide multiple datasets, with only one precise, known, setting modified between each variation.
Finally, customization capabilities of \nameCyberRange{} allow replication and update of existing datasets.

This Section presents \nameCyberRange{}'s main components, how they interact and how they allow to customize the datasets produced. 
It presents the customizable settings of the legitimate and malicious traffic contained in the datasets generated by \nameCyberRange{}.
Finally, the benefits of \nameCyberRange{} are not solely about customization. 
This Section's end presents extra improvements fixing performance and realism issues. 

\subsection{Architecture}\label{subsec:archi}

\begin{figure*}[htbp]
    \includegraphics[width=0.95\textwidth]{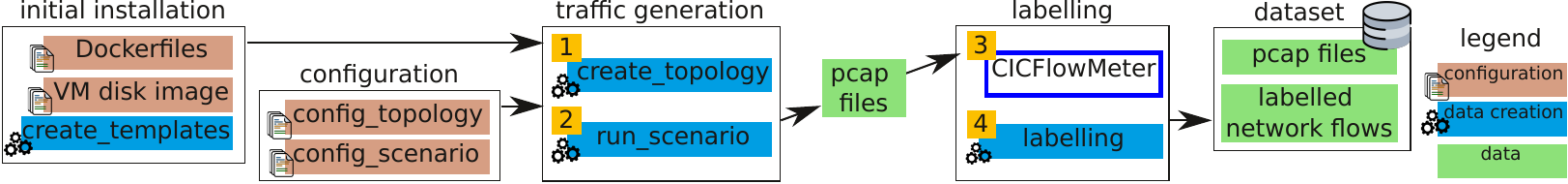}
    \caption{\nameCyberRange{}'s workflow}
    \label{fig:workflow}
\end{figure*}

\begin{figure}[htbp]
    \includegraphics[width=0.47\textwidth]{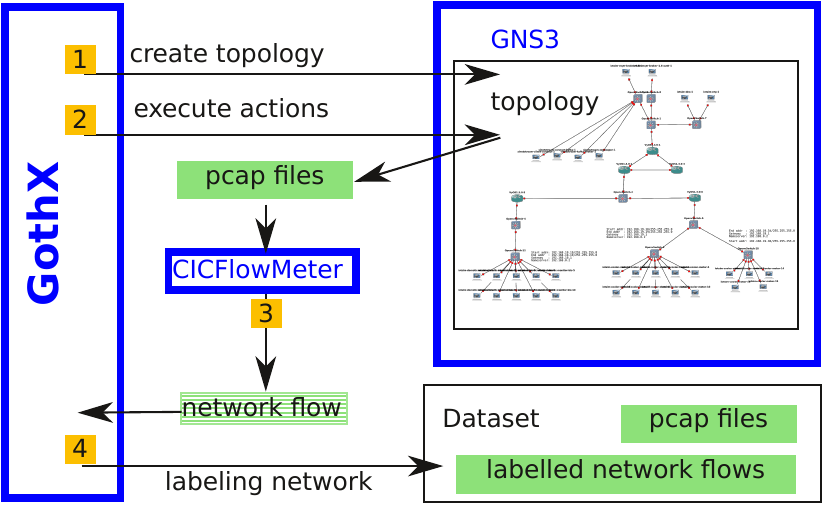}
    \caption{\nameCyberRange{}'s interaction with other tools}
    \label{fig:gns3_interact}
\end{figure}

To generate customized labeled datasets, \nameCyberRange{} follows the workflow depicted in  Figure~\ref{fig:workflow}.
At first, it is necessary to create templates of the future topology's nodes.
Then, \nameCyberRange{}'s scripts use configuration files to influence the data they generate. 
In the end, this data is labeled to form the final dataset.
Figure~\ref{fig:gns3_interact} explains how \nameCyberRange{} interacts with GNS3 and leverages CICFlowMeter~\cite{cicflowmeter}. 
Basically, \nameCyberRange{}'s role is to automatically generate a topology in GNS3 and execute actions in this topology. 
CICFlowMeter is used to extract network flows.
\nameCyberRange{}'s GitHub repository provides documentation about installation and usage to create the custom topology and run a configured scenario as well as the labeling process.

The configuration file \texttt{config\_topology} defines different topology's configurations.
A configuration influences how many nodes of each type will be generated in the topology.
The second configuration file \texttt{config\_scenario} references a previously created topology.
The scenario defines which nodes will be started and how they will behave.
It also describes which network traffic will be captured.

\texttt{create\_topology} creates a topology in GNS3 based on the topology's configuration file (step 1 on Figure~\ref{fig:gns3_interact}).
Then, \texttt{run\_scenario} uses \texttt{config\_scenario} to automatically perform legitimate and malicious actions in the topology (step 2 on Figure~\ref{fig:gns3_interact}).
It creates some \texttt{pcap} files containing raw network packets generated during the scenario execution. 

Labeling is required to allow usage of the dataset for supervised machine learning because \texttt{pcap} files mix legitimate and malicious traffic.
The labeling is a two-steps process. 
At first (step 3 on Figure~\ref{fig:gns3_interact}), CICFlowMeter extracts, in a \texttt{csv} file, network flows from each \texttt{pcap} file, alongside different features.
Then a labeling script, part of \nameCyberRange{}, automatically labels each network flow (step 4 on Figure~\ref{fig:workflow}). 
It allows anyone to distinguish legitimate traffic among different types of malicious traffic.
For example, this is useful in the case of training a model performing multi-class classification to differentiate each step of a long attack like discriminating DDoS from ports scan.

\subsection{Customization} \label{subsec:customization}

\begin{table}[htpb]
    \caption{Customizable topology and scenario parameters}
    \label{tab:customizable-parameters}
    \begin{tabular}{|c|c|}
        \hline
        \multicolumn{1}{|l|}{\textbf{Legitimate traffic}}  & \multicolumn{1}{l|}{\textbf{Malicious traffic}} \\ \hline
        \multicolumn{1}{|l|}{Sensors count}                & \multicolumn{1}{l|}{Parameters of attack tool}  \\ \hline
        \multicolumn{1}{|l|}{Messages rate*}              & \multicolumn{1}{l|}{Strength of DDoS attack}    \\
        \multicolumn{1}{|l|}{(periodic/random)}           & \multicolumn{1}{l|}{(e.g. payload size)}                           \\ \hline
        \multicolumn{1}{|l|}{(In)activity duration*}       & \multicolumn{1}{l|}{\% of compromised sensors}  \\ \hline
        \multicolumn{1}{|l|}{Which data, from a dataset}   & \multicolumn{1}{l|}{Sleep time}                 \\
        \multicolumn{1}{|l|}{of real sensors, is sent*}    & \multicolumn{1}{l|}{between attack steps}       \\ \hline
        \multicolumn{1}{|l|}{Traffic volume (MQTT/Kafka)*} & \multicolumn{1}{l|}{}                           \\ \hline
    \end{tabular}
    *customizable for each sensor independently
\end{table}

With \nameCyberRange{}, both the topology and the scenario can be easily customized using the configuration files described in Section~\ref{subsec:archi}.
It is very helpful to generate datasets dedicated to the need.
Customization of the scenario is useful in the context of anomaly detection to analyze the efficiency of the detector when legitimate traffic varies but the attack is the same, or vice versa.
For example, it can be useful in AI and XAI because it helps to study the impact of a specific parameter on a machine learning model for anomaly detection.

Table~\ref{tab:customizable-parameters} details parameters that can be customized concerning the malicious and legitimate traffic generated by \nameCyberRange{}.
The malicious traffic comes from the automatic execution of an attack scenario.
This scenario is implemented and published jointly with \nameCyberRange{}.
Modifying the parameters, one can define a waiting time of three hours between payload transfer and DDoS start, for example.
Varying the arguments of the attack tools will impact the strength, speed and therefore stealthiness of the attack.
Attackers tend to increase their stealthiness by increasing the time between attack stages or by making slower scans or brute-force.
\nameCyberRange{}'s scenario configuration file allows anyone to study the impact of such variations on the ability to detect the attack.

\subsection{Solved performances and realism issues}

\nameCyberRange{} includes various improvements and extra functionalities to state-of-the-art as detailed below.

No graphical user interface is necessary to use \nameCyberRange{}, which enables its usage on a remote server.
Embedded install scripts allow to install and setup all requirements.
Similarly, usage only consists of optionally modifying configuration files and executing scripts.
It is therefore possible to go from an empty computer to the full execution of a scenario using a command line interface only.

When deploying a topology in GNS3, configuration of VM nodes (i.e. install image disk, set network configuration) is time-consuming.
With \nameCyberRange{}, all VM nodes are configured in parallel and in the background.
It significantly reduces the time required to deploy a topology, making it more convenient to quickly experiment with new topologies.

In \nameCyberRange{}, the realism of the data contained in MQTT messages is assured by reading line by line publicly available datasets containing data collected from real sensors.
Furthermore, it is possible to choose which columns of the dataset is read by a simulated sensor.
This way, multiple simulated sensors can read from the same dataset and still send different data, which improves data diversity.
Moreover, this new feature allows the definition of network packet size by modifying the number of columns contained in the published MQTT message.

%% file: mqttset.tex
\nameCyberRange{} allows to replicate MQTTset~\cite{vaccari_mqttset_2020}, a public dataset containing network packets involving MQTT.
This validates \nameCyberRange{}'s ability to generate good-quality data. 
It also demonstrates that thanks to customization, \nameCyberRange{} can generate data following precise characteristics previously defined. 
Indeed, we could check that our dataset presents the same characteristics as those of MQTTset. 
For example, we have the same number of packets/sec. and Bits/sec. in TCP flows.

MQTTset was produced to train machine learning models to detect security issues in MQTT exchanges.
It is composed of six IoT traffic traces.
One trace contains only legitimate traffic.
It represents 10 distinct IP addresses publishing messages to 1 MQTT broker.
5 IP addresses send data periodically, every 60, 120 or 180 seconds.
5 IP addresses send data at a random interval, with a mean time of one second between two messages.
This file was generated using the traffic generator IoT-Flock~\cite{ghazanfar_iot-flock_2020}.
This tool generates a \texttt{pcap} file with network packets as they would appear in real communications between IoT devices.
Each of the five other traffic traces in MQTTset contains network packets about a different type of attack against MQTT.
Table~\ref{tab:mqttset_attacks} details the type of attack and the tool used to generate the file.

\begin{table}[htbp]
    \centering
    \caption{Attack types and corresponding tools in MQTTset}
    \label{tab:mqttset_attacks}
    \begin{tabular}{|l|l|}
        \hline
        \textbf{Attack type}                & \textbf{Tool}            \\ \hline
        MQTT publish flood, CVE-2018-1684   & IoT-Flock~\cite{ghazanfar_iot-flock_2020} \\ \hline
        Flood DoS                           & MQTT-malaria~\cite{mqtt_malaria_2024} \\ \hline
        SlowITe                             & SlowTT~\cite{vaccari_slowite_2020}    \\ \hline
        Malformed data                     & MQTTSA~\cite{palmieri_mqttsa_2019}       \\ \hline
        Authentication bruteforce           & MQTTSA~\cite{palmieri_mqttsa_2019}       \\ \hline
    \end{tabular}
\end{table}

With \nameCyberRange{}, we produced a dataset of six traffic traces with network traffic having the same characteristics as MQTTset.
Files containing an attack were replicated using the same attack tools, except for the one generated by IoT-Flock (first line in Table~\ref{tab:mqttset_attacks}).
The latter was replicated using MQTT-malaria~\cite{mqtt_malaria_2024}, a tool for testing scalability in MQTT environments, with some arguments leading to the same effect.
(namely, \texttt{malaria publish -P 1 -n 265 -H 192.168.2.1 -s 30700}).
Each of the six traffic traces in our dataset has the same number of distinct IP addresses as in their corresponding traffic trace in MQTTset.
We confirmed that our datasets have similar statistics in each TCP conversation regarding the number of packets, the number of bytes, the bits rate (bits/sec.).
For example, the flood DoS (second line in Table~\ref{tab:mqttset_attacks}) in MQTTset consist of 44M packets received by an MQTT broker at the bit rate of 2.8K bits/sec. during 126 seconds.
In our dataset, there are 62M packets received at the bit rate of 4.5K bits/sec. during 110 seconds.
Finally, we could confirm that ML models for anomaly detection had similar results on MQTTset and our replication. 

Our dataset contains some differences with MQTTset, which makes it better than the original dataset.
Our legitimate traffic was generated in \nameCyberRange{} with one node being the MQTT broker and ten nodes the sensors.
Real MQTT messages are exchanged between the broker and the sensors.
In MQTTset on the contrary, legitimate traffic was synthesized with IoT-Flock, which only crafts network packets.
Real interactions allow anyone to study how the attack interferes with legitimate actions.
This is particularly interesting for DoS which goal is precisely to disrupt legitimate usage.
In MQTTset, legitimate traffic is fully separated from the malicious one while \nameCyberRange{} can mix both.
A mix is more realistic because real network captures of an attack usually also contain legitimate traffic and the main difficulty of intrusion detection systems (IDS) is to distinguish normal and malicious traffic.
Furthermore, our dataset is more realistic than MQTTset because it is more diverse. 
It contains not only network data involving  MQTT, but also DNS, NTP, ARP and ICMP exchanges.
This is the diversity that exists in real network traffic and IDS must be trained to deal with it to be efficient in real networks.

%% file: attack_scenario.tex
\subsection{New realistic complete attack scenario}\label{subsec:attack}

As a second use case, we used \nameCyberRange{} to generate a dataset from the automatic execution of a scenario mixing legitimate and malicious actions between IoT devices.
This section presents the topology where the scenario takes place.
It details the different steps of the complete \textit{kill-chain} followed by an attacker, from the initial compromision by exploiting a recent critical vulnerability (CVE-2023-25194~\cite{cve25194}) to a DDoS.
Analysis of the generated network traffic validates the smooth execution of the scenario, both for legitimate and malicious actions.

\subsubsection{Topology}
\begin{figure}
    \centering
    \includegraphics[width=0.3\textwidth]{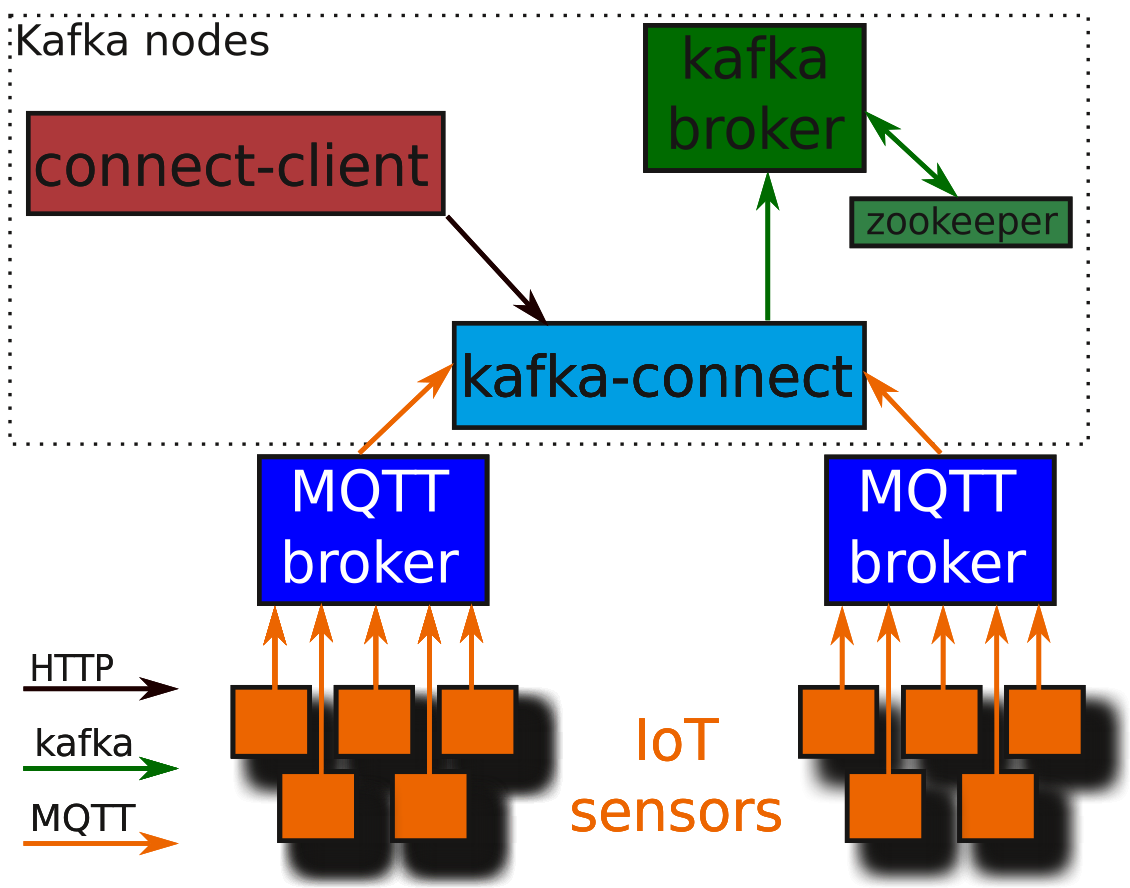}
    \caption{Overview of the simulated network}
    \label{fig:topo_sinet}
\end{figure}
The topology used in the scenario at the origin of our dataset is depicted in Figure~\ref{fig:topo_sinet}.
It contains:
\begin{itemize}
    \item 450 nodes simulating IoT sensors sending MQTT messages
    \item 3 MQTT brokers
    \item 1 Kafka broker
    \item 1 node <<\textit{kafka-connect}>>, executing Kafka-connect
    \item 1 node <<\textit{client-connect}>>, the client of <<\textit{kafka-connect}>>
\end{itemize}
Using the topology configuration file described in Section~\ref{subsec:archi}, a user of \nameCyberRange{} who wants to perform a small-scale experiment can easily generate a smaller topology with fewer than 450 IoT devices or fewer than three MQTT brokers or no node related to Kafka.

Legitimate traffic is generated by IoT sensors sending messages to MQTT brokers.
Then, Kafka-connect is used to read topics in MQTT brokers and transfer the messages to a Kafka broker.
The simulated environment is similar to existing IoT architectures mixing MQTT and Kafka traffic, like SINETStream~\cite{sinetstream2021, takefusa_sinetstream_2021}.
Two MQTT brokers receive clear text messages, with one of them requiring device authentication with a login and password before publishing.
The third receives TLS encrypted messages, without prior authentication.

Some sensitive, but plausible, settings have been defined to introduce vulnerabilities allowing execution of a realistic attack scenario from the initial compromission of a node to a DDoS by multiple IoT sensors.
On IoT devices, the package \texttt{ssh-server} is available so that it's possible to activate a remote SSH connection.
Also, the Kafka-connect component runs version 7.3.1 from confluent.
This version was released in December 2022 and is the last version vulnerable to the CVE-2023-25194~\cite{cve25194}.
Moreover, Kafka-connect is a Java program for which we intentionally set the system property \texttt{UnableUnsafeSerialization} to \texttt{true}.
Therefore, an attacker able to send \texttt{POST} requests to Kafka-connect can exploit the CVE-2023-25194 and get a remote code execution (RCE) on the node executing Kafka-connect.

\subsubsection{Steps of complete attack scenario}

Based on what can be seen in recent attacks launched against IoT networks~\cite{haseeb_iot_killchain_2020, rachit_security_iot_trends_2021, abbasi_security_threat_2022}, we introduce a new realistic attack scenario.
Our scenario considers an internal threat with the attacker initially controlling the node <<\textit{client-connect}>>.
He begins by sending legitimate \texttt{POST} requests to <<\textit{kafka-connect}>> and then starts executing malicious actions remotely. The scenario is divided into six steps.
\begin{enumerate}
    \item From the node <<\textit{client-connect}>>, the attacker exploits the CVE-2023-25194 affecting Kafka-connect.
    It allows him to gain a remote code execution (RCE) on the node <<\textit{kafka-connect}>>.
    \item The attacker uses the RCE to transfer attack tools from <<\textit{client-connect}>> to <<\textit{kafka-connect}>> with the utility \texttt{wget}.
    Then, he configures the attack tools on <<\textit{kafka-connect}>> and he opens a reverse shell using \texttt{netcat}.
    \item From <<\textit{kafka-connect}>>, the attacker scans the network, searching for devices with port 22 listening, which is the usual port for SSH.
    The command executed is \texttt{./nmap -Pn -oG ips.txt 192.168.18-20.10-150 --max-rate 0.7 -p 22}.
    \item For nodes detected as listening to SSH communication, the attacker searches for the right SSH credentials by trying multiple combinations of username/password using dictionaries of commonly used credentials.
    This is done from <<\textit{kafka-connect}>>, with the command \texttt{hydra/hydra -o success.txt -M ssh\_ips.txt ssh -f -L u.txt -P p.txt -t 2} executed.
    \item For each node where he found the right SSH credentials, the attacker transfers his payload from <<\textit{kafka-connect}>> to the compromised node using \texttt{scp}, which is dedicated to transferring files over SSH.
    Here, the payload is the tool \texttt{mqttsa}.
    \item From <<\textit{kafka-connect}>> and using SSH, the payload is triggered simultaneously on all compromised nodes which all execute the command \texttt{mqttsa -fc 100 -fcsize 10 -sc 2400 192.168.2.1}.
    This is the beginning of the DDoS targeting the MQTT broker at the IP address \texttt{192.168.2.1}.
\end{enumerate}
When a user of \nameCyberRange{} wants to execute this scenario, he can choose the arguments of attack commands with the scenario configuration file presented in Section~\ref{subsec:archi}.
Moreover, a user only interested in studying the DDoS can skip all steps except for the last one.
In this case, desired nodes, simulated by docker-containers, are considered as already compromised.
\texttt{mqttsa} is installed on the compromised nodes by copying the file from the machine hosting \nameCyberRange{} to the containers, without generating network traffic.
Similarly, the payload is triggered by executing a command on the containers from the host, without generating network traffic.
This use case with DDoS only was used to evaluate in-kernel anomaly detection using eBPF~\cite{ebpf_mirai} with various proportions of compromised nodes.

\subsubsection{Validation of generated network traffic}

\begin{figure}[htpb]
    \centering
    \includegraphics[width=0.47\textwidth]{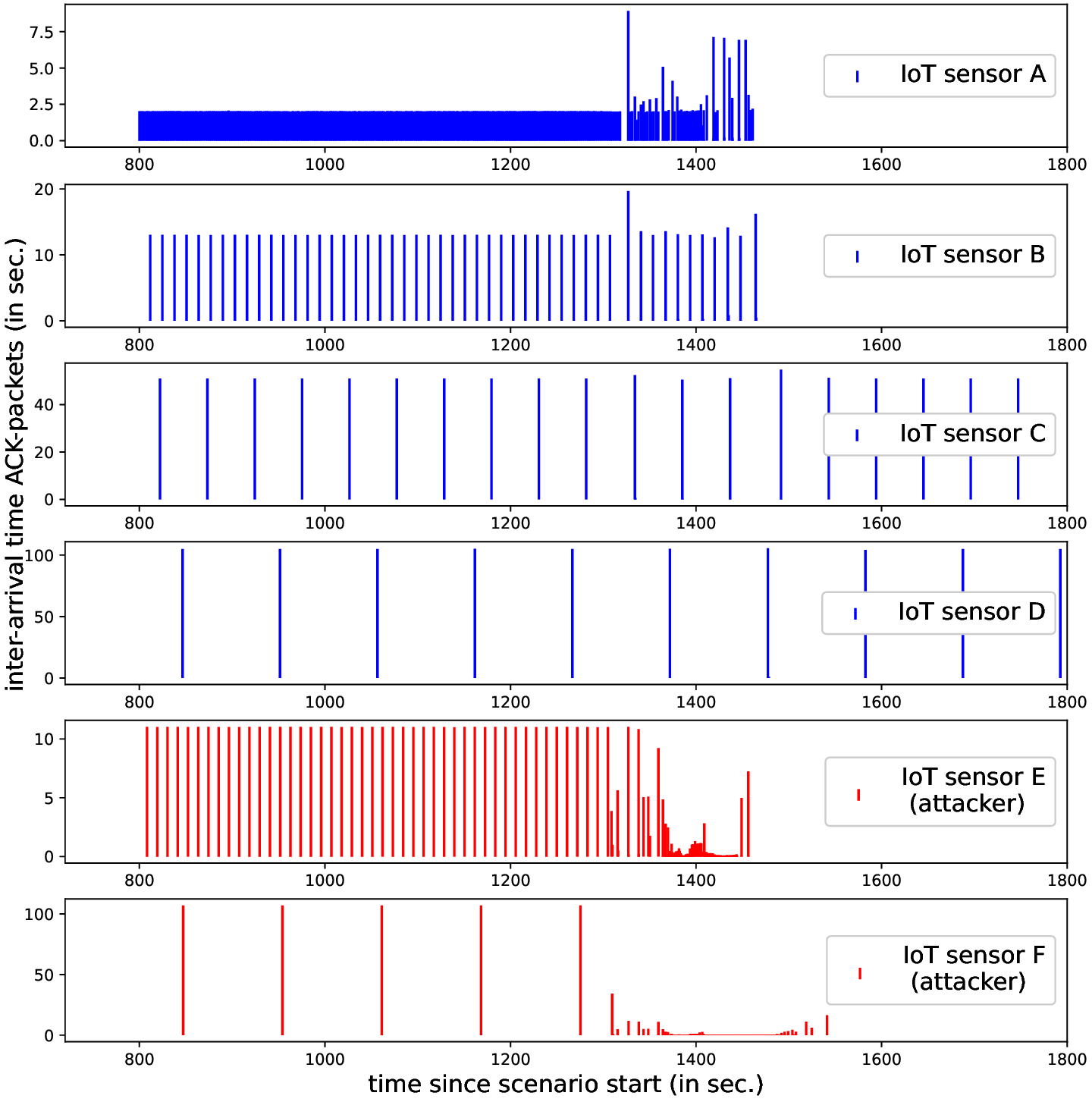}
    \caption{Inter-arrival time of ACK-packets during scenario}
    \label{fig:evaluation}
\end{figure}

To validate that the generated traffic was the one expected, we analyzed the traffic traces resulting from the execution of our scenario.
In Figure~\ref{fig:evaluation}, each plot shows, for a given destination IoT sensor, the inter-arrival time of some specific packets with respect to the time since the scenario's beginning in seconds.
We focus on packets emitted by MQTT brokers sending acknowledgments to IoT sensors, the ones with the TCP flag ACK.
A first ACK-packet is sent when an IoT sensor establishes an MQTT connection.
Almost immediately after (the inter-arrival time is near zero), a second ACK-packet is sent when the broker correctly receives an MQTT message.
These plots allow to verify that (before being disrupted by the attack) MQTT brokers correctly receive messages at the rate defined in the file \texttt{config\_scenario}, even when some sensors publish messages faster than one message per second.

When everything works fine, the cycle repeats itself periodically.
This can be seen on the left of the plots in Figure~\ref{fig:evaluation}, before 1200 seconds.
For example, the second plot shows an MQTT broker acknowledging MQTT connections and publications every 12 seconds for the IoT sensor B.

During the DDoS, MQTT brokers acknowledge, to the nodes launching the attack, many MQTT connections and publications in a very short time.
The inter-arrival time between two ACK-packets sent to attacking nodes is very small.
This is represented after about 1200 seconds of execution, by the drop near zero of plots E and F related to the nodes representing compromised IoT sensors responsible for the DDoS.
For non-attacking nodes, acknowledgments of some MQTT connections are missed or delayed.
This is shown by the increase of plots A and B, related to non-attacking IoT sensors.
This phenomenon is more noticeable for nodes with a high publication frequency (e.g.\ IoT sensor A) but appears for almost all nodes.

We notice that after DDoS, most plots stop before the end of the scenario.
IoT sensors were unable to contact their MQTT broker.
They considered the broker as unreachable and stopped sending connection requests.
Before the DDoS, 225 IoT nodes received ACK-packets from the broker targeted by the attack.
The scenario last about 1800 seconds and after 1700 seconds, the victim MQTT broker stopped sending ACK-packet to 82\% of these IoT nodes.
This analysis shows that the DDoS is efficient because it definitely disrupts the normal behavior of nodes.

\subsubsection{Provided dataset based on our attack scenario}

The dataset produced is composed of a text file detailing settings of the scenario execution, three \texttt{pcap} files containing raw network packets, associated with three label files.
The text file gives, among others, the commands executed by the attacker at each attack step and the IP address of nodes involved in the DDoS.

Our goal was to have the minimum number of network probes to capture all the network traffic generated during the scenario.
It leads to three probes recording three traffic traces in three \texttt{pcap} files.
One \texttt{pcap} file contains only legitimate traffic about the communication between Zookeeper and the Kafka broker.
Another \texttt{pcap} file contains all traffic related to the node <<\textit{kafka-connect}>>.
In particular, there is the legitimate transfer of messages from MQTT brokers to the Kafka broker.
This is mixed with the malicious traffic related to the exploitation of the CVE~\cite{cve25194}, the attack tool transfer on node <<\textit{kafka-connect}>> and the commands sent through the \texttt{netcat} reverse shell.
The third and last \texttt{pcap} file contains all legitimate traffic about IoT sensors sending MQTT messages to MQTT brokers.
This is mixed with the following malicious activity: ports scan, credential brute force, payload transfer on compromised devices via \texttt{scp} and DDoS of the MQTT broker.

Since \texttt{pcap} files mix legitimate and malicious traffic, labeling is required to allow the usage of the dataset for supervised machine learning.
Furthermore, our labeling is useful to distinguish different types of malicious traffic in case of training an IDS to differentiate each step of the attack like discriminating DDoS from ports scan for example.
It is possible to know, for every single packet, whether it corresponds to benign traffic (labeled as \textit{normal}) or if it is part of the attack process.
Labels about the attack are \textit{cve\_exploitation}, \textit{reverse\_shell}, \textit{scan\_ports}, \textit{credentials\_bruteforce}, \textit{transfer\_payload\_to\_iot} and \textit{mqttsa\_slowite}.
They detail to which attack step the packet is related to.
Our labeling script is adaptive to the various settings applied when the scenario was executed.
Using \nameCyberRange{}, everyone can generate his custom dataset fitting his need and label it using our labeling script.

\nameCyberRange{}'s customization capabilities, detailed in Table~\ref{tab:customizable-parameters}, allow anyone to generate variations of our dataset.
For example, one can easily change one line in the scenario configuration file to consider a DDoS with 25\% of compromised nodes, instead of 50\%.
With this new dataset, he can see if the disruption of legitimate traffic is reduced and if an IDS can still detect the attack.
Another interesting experiment would be to evaluate the evolution of an IDS performance when the proportion of legitimate over malicious traffic increases.
\nameCyberRange{}'s customization makes this possible by allowing an easy variation in the size and frequency of legitimate messages while keeping the same attack.

%% file: scalable_repro.tex
\subsection{Identification of the scalability bottleneck}

The scalability of \nameCyberRange{} is defined as the number of simulated devices that can easily run simultaneously.
To be the most realistic possible, scalability is required because real IoT networks and DDoS involve many devices.
We want to easily increase the number of IoT sensors in the topology.
In particular, we identified the bottleneck to the maximum number of IoT sensors that can be simulated in \nameCyberRange{}.
We assessed two challenges to overcome to achieve scalability: hardware resources and data realism.
Finally, we measured execution time. 

\paragraph{Hardware resources}
In terms of memory (RAM) consumption, Gotham and \nameCyberRange{} behave the same way.
Based on measurements of Gotham's paper, we can infer that $R$, the amount of RAM in megabytes necessary to run nodes, is about
$R=q * 470 + d * 37$, where $d$, respectively $q$, is the number of nodes based on docker, respectively based on qemu VMs, running simultaneously.
In \nameCyberRange{}, we could execute a scenario involving 450 IoT devices.
Such topology is made of 498 docker nodes for IoT devices, switches, brokers, and four qemu VMs nodes for routers.
It uses 20.4GB of RAM, which is not a strong limitation, even for some laptops. 

Another hardware resource we considered that could limit scalability is the CPU usage.
Our measurements showed that the number of running nodes does not strongly impact CPU usage.
However, depending on what a node is doing, it may use a lot of CPU power.
In particular, a scenario with some settings leading to a strong DDoS creates many processes on the machine running the topology and CPU consumption is high.
Even with only two IoT devices set to launch the DDoS, this can lead a laptop to freeze if settings are dis-proportioned.
It can be noted that, since \nameCyberRange{} adds the ability to be executed on a remote server, it eases access to computers with a lot of memory and CPU power.

\paragraph{Realism}
The realism of the simulated network must be kept in mind while increasing the number of devices.
In particular, duplicating dozens of times devices with the same behavior easily increases the number of simulated nodes but is only minimally representative of a real IoT network.
\nameCyberRange{} realism lies in its ability to generate network traffic similar to the one in a real IoT network with multiple different sensors sending their data to some brokers.
Data sent by IoT devices simulated in \nameCyberRange{} is read from public datasets where each column contains data registered by a real sensor.
The cumulated number of real sensors represented in the public datasets we use does not reach 450.
Therefore, with 450 simulated sensors in the topology, not all simulated sensors can publish data from a different real sensor.
We lose realism with multiple simulated sensors publishing the same data from the same real sensor.
However, configuring simulated sensors to publish different combinations of data from multiple real sensors mitigates the realism loss.

\paragraph{Execution time}
Two different execution times can be distinguished.
The execution time for data generation is the same as the scenario duration. This time is fully customizable using the scenario configuration file and depends on the temporal scope of the dataset desired by \nameCyberRange{}'s user.
Concerning the time needed to deploy a topology, prior to a scenario execution, it depends on the number of VM nodes and docker nodes in the topology but remains reasonable.
Docker nodes are fast to configure.
Creating and configuring one VM takes a median time of 278 seconds based on our multiple measurements. 
Thanks to the parallel configuration of VMs in the background, multiple VMs can be configured simultaneously and during the creation of docker nodes.
Following our multiple executions on a laptop, \nameCyberRange{} takes about 26 minutes to create a large topology composed of four VM nodes and 498 docker nodes. 
The topology we used to replicate MQTTset contains 5 VM nodes and 29 docker nodes and is created in about less than 3 minutes.

\subsection{Replicability}
Replicability is the recreation of the same experimental apparatus, and using it to perform exactly the same experiment.
We paid particular attention to replicability at different levels.

At first, the traffic generator itself is replicable.
The code is open-source, and available on GitHub.
We checked that \nameCyberRange{}'s documentation allowed peers to use our traffic generator without external help.
Validation of the portability was done by verifying \nameCyberRange{}'s ability to be installed and generate the same topology on different laptops and on a remote server with the operating
systems Fedora and Ubuntu.
This verifies that \nameCyberRange{} can be used on different host computers and is not dependent on a specific operating system.

Secondly, we assessed the replicability of the generated dataset.
We could verify that executing \nameCyberRange{} twice with the same settings generates the same dataset in the end.
One of \nameCyberRange{}'s strengths is its ability to replicate existing datasets.
Specifically, users can take a public dataset and generate a new one with similar characteristics. 
This is one of the use cases detailed in Section~\ref{subsec:mqttset} for MQTTset.
Furthermore, \nameCyberRange{} is useful for generating variations of an existing dataset.
A user can replicate a famous dataset and make sure he gets similar results with the copy.
Then he can make new experiments to understand what would have changed if a given setting had been different in the original dataset.

%% file: conclu.tex
GothX is an open-source, replicable and highly customizable traffic generator for heterogeneous IoT networks. 
It sets up virtual devices, network topologies and specific simulation scenarios in a completely automatic fashion. Parameters of simulation are easily modifiable through configuration files that are well documented in the GitHub repository. From its diverse simulation scenarios, GothX captures rich mixed IoT traffic and creates traffic datasets automatically. The following new features in the state-of-the-art of traffic generation are implemented in GothX: highly customizable node behavior, specific mechanisms of traffic diversification for both legitimate and malicious traffic, automatic labeling of generated dataset, and increased heterogeneity of network topology through automatic configuration of MQTT and Kafka networks.

We validated GothX with two use cases: the replication and enrichment of the simulation that generated the mixed traffic dataset MQTTset, emphasizing the diversification of legitimate traffic parameters; and the simulation of a six-step scenario involving CVE-2023-25194 exploit, nodes infection, and the subsequent DDoS attack, emphasizing the diversification of malicious traffic parameters. Through these two use cases, Gotham generated the mixed IoT traffic used to create two new datasets, both of which we make available publicly. The current scalability of the solution allows up to 450 nodes simulated at a cost of 20.4GB of RAM. CPU cost is also feasible for single machines even during the simulation of weaker DDoS attacks.

%% file: paper.bbl
%%% -*-BibTeX-*-
%%% Do NOT edit. File created by BibTeX with style
%%% ACM-Reference-Format-Journals [18-Jan-2012].

\begin{thebibliography}{33}

%%% ====================================================================
%%% NOTE TO THE USER: you can override these defaults by providing
%%% customized versions of any of these macros before the \bibliography
%%% command.  Each of them MUST provide its own final punctuation,
%%% except for \shownote{}, \showDOI{}, and \showURL{}.  The latter two
%%% do not use final punctuation, in order to avoid confusing it with
%%% the Web address.
%%%
%%% To suppress output of a particular field, define its macro to expand
%%% to an empty string, or better, \unskip, like this:
%%%
%%% \newcommand{\showDOI}[1]{\unskip}   % LaTeX syntax
%%%
%%% \def \showDOI #1{\unskip}           % plain TeX syntax
%%%
%%% ====================================================================

\ifx \showCODEN    \undefined \def \showCODEN     #1{\unskip}     \fi
\ifx \showDOI      \undefined \def \showDOI       #1{#1}\fi
\ifx \showISBNx    \undefined \def \showISBNx     #1{\unskip}     \fi
\ifx \showISBNxiii \undefined \def \showISBNxiii  #1{\unskip}     \fi
\ifx \showISSN     \undefined \def \showISSN      #1{\unskip}     \fi
\ifx \showLCCN     \undefined \def \showLCCN      #1{\unskip}     \fi
\ifx \shownote     \undefined \def \shownote      #1{#1}          \fi
\ifx \showarticletitle \undefined \def \showarticletitle #1{#1}   \fi
\ifx \showURL      \undefined \def \showURL       {\relax}        \fi
% The following commands are used for tagged output and should be
% invisible to TeX
\providecommand\bibfield[2]{#2}
\providecommand\bibinfo[2]{#2}
\providecommand\natexlab[1]{#1}
\providecommand\showeprint[2][]{arXiv:#2}

\bibitem[mqt(2024)]%
        {mqtt_malaria_2024}
 \bibinfo{year}{2024}\natexlab{}.
\newblock \bibinfo{title}{etactica/mqtt-malaria}.
\newblock
\newblock
\urldef\tempurl%
\url{https://github.com/etactica/mqtt-malaria}
\showURL{%
\tempurl}
\newblock
\shownote{(Accessed on May 13th, 2024)}.


\bibitem[cve(2024)]%
        {cve25194}
 \bibinfo{year}{2024}\natexlab{}.
\newblock \bibinfo{title}{{NVD} - {CVE}-2023-25194}.
\newblock
\newblock
\urldef\tempurl%
\url{https://nvd.nist.gov/vuln/detail/CVE-2023-25194}
\showURL{%
\tempurl}
\newblock
\shownote{(Accessed on May 13th, 2024)}.


\bibitem[Abbasi et~al\mbox{.}(2022)]%
        {abbasi_security_threat_2022}
\bibfield{author}{\bibinfo{person}{Mahmoud Abbasi}, \bibinfo{person}{Marta
  Plaza-Hernandez}, \bibinfo{person}{Javier Prieto}, {and}
  \bibinfo{person}{Juan~M. Corchado}.} \bibinfo{year}{2022}\natexlab{}.
\newblock \showarticletitle{Security in the {Internet} of {Things}
  {Application} {Layer}: {Requirements}, {Threats}, and {Solutions}}.
\newblock \bibinfo{journal}{\emph{IEEE Access}}  \bibinfo{volume}{10}
  (\bibinfo{year}{2022}), \bibinfo{pages}{97197--97216}.
\newblock
\showISSN{2169-3536}
\urldef\tempurl%
\url{https://doi.org/10.1109/ACCESS.2022.3205351}
\showDOI{\tempurl}


\bibitem[Anande and Leeson(2023)]%
        {anande2023}
\bibfield{author}{\bibinfo{person}{Tertsegha Anande} {and}
  \bibinfo{person}{Mark Leeson}.} \bibinfo{year}{2023}\natexlab{}.
\newblock
  \showarticletitle{WRAP-Generative-adversarial-networks-GANs-survey-network-traffic-generation-2022}.
\newblock \bibinfo{journal}{\emph{International Journal of Machine Learning and
  Computing}}  \bibinfo{volume}{12} (\bibinfo{date}{10} \bibinfo{year}{2023}),
  \bibinfo{pages}{333 -- 343}.
\newblock
\urldef\tempurl%
\url{https://doi.org/10.18178/ijmlc.2022.12.6.1120}
\showDOI{\tempurl}


\bibitem[Archive(1999)]%
        {kdd99}
\bibfield{author}{\bibinfo{person}{The UCI~KDD Archive}.}
  \bibinfo{year}{1999}\natexlab{}.
\newblock \bibinfo{title}{{ KDD Cup 1999 Data}}.
\newblock
\newblock
\urldef\tempurl%
\url{https://kdd.ics.uci.edu/databases/kddcup99/kddcup99.html}
\showURL{%
\tempurl}
\newblock
\shownote{(Accessed on May 13th, 2024)}.


\bibitem[Asplund and Nadjm-Tehrani(2016)]%
        {asplund2016}
\bibfield{author}{\bibinfo{person}{Mikael Asplund} {and} \bibinfo{person}{Simin
  Nadjm-Tehrani}.} \bibinfo{year}{2016}\natexlab{}.
\newblock \showarticletitle{Attitudes and Perceptions of {IoT} Security in
  Critical Societal Services}.
\newblock \bibinfo{journal}{\emph{IEEE Access}}  \bibinfo{volume}{4}
  (\bibinfo{year}{2016}), \bibinfo{pages}{2130--2138}.
\newblock
\urldef\tempurl%
\url{https://doi.org/10.1109/ACCESS.2016.2560919}
\showDOI{\tempurl}


\bibitem[Bagchi et~al\mbox{.}(2020)]%
        {bagchi2020}
\bibfield{author}{\bibinfo{person}{Saurabh Bagchi}, \bibinfo{person}{Tarek~F.
  Abdelzaher}, \bibinfo{person}{Ramesh Govindan}, \bibinfo{person}{Prashant
  Shenoy}, \bibinfo{person}{Akanksha Atrey}, \bibinfo{person}{Pradipta Ghosh},
  {and} \bibinfo{person}{Ran Xu}.} \bibinfo{year}{2020}\natexlab{}.
\newblock \showarticletitle{New Frontiers in {IoT}: Networking, Systems,
  Reliability, and Security Challenges}.
\newblock \bibinfo{journal}{\emph{IEEE Internet of Things Journal}}
  \bibinfo{volume}{7}, \bibinfo{number}{12} (\bibinfo{year}{2020}),
  \bibinfo{pages}{11330--11346}.
\newblock
\urldef\tempurl%
\url{https://doi.org/10.1109/JIoT.2020.3007690}
\showDOI{\tempurl}


\bibitem[Bounceur et~al\mbox{.}(2018)]%
        {bounceur_cupcarbonlab_2018}
\bibfield{author}{\bibinfo{person}{Ahcene Bounceur}, \bibinfo{person}{Olivier
  Marc}, \bibinfo{person}{Massinissa Lounis}, \bibinfo{person}{Julien Soler},
  \bibinfo{person}{Laurent Clavier}, \bibinfo{person}{Pierre Combeau},
  \bibinfo{person}{Rodolphe Vauzelle}, \bibinfo{person}{Loic Lagadec},
  \bibinfo{person}{Reinhardt Euler}, \bibinfo{person}{Madani Bezoui}, {and}
  \bibinfo{person}{Pietro Manzoni}.} \bibinfo{year}{2018}\natexlab{}.
\newblock \showarticletitle{{CupCarbon}-{Lab}: {An} {IoT} emulator}. In
  \bibinfo{booktitle}{\emph{2018 15th {IEEE} {Annual} {Consumer}
  {Communications} \& {Networking} {Conference} ({CCNC})}}.
  \bibinfo{publisher}{IEEE}, \bibinfo{address}{Las Vegas, NV},
  \bibinfo{pages}{1--2}.
\newblock
\showISBNx{978-1-5386-4790-5}
\urldef\tempurl%
\url{https://doi.org/10.1109/CCNC.2018.8319313}
\showDOI{\tempurl}


\bibitem[De~Keersmaeker et~al\mbox{.}(2023)]%
        {keersmaeker2023}
\bibfield{author}{\bibinfo{person}{François De~Keersmaeker},
  \bibinfo{person}{Yinan Cao}, \bibinfo{person}{Gorby~Kabasele Ndonda}, {and}
  \bibinfo{person}{Ramin Sadre}.} \bibinfo{year}{2023}\natexlab{}.
\newblock \showarticletitle{A Survey of Public IoT Datasets for Network
  Security Research}.
\newblock \bibinfo{journal}{\emph{IEEE Communications Surveys \& Tutorials}}
  \bibinfo{volume}{25}, \bibinfo{number}{3} (\bibinfo{year}{2023}),
  \bibinfo{pages}{1808--1840}.
\newblock
\urldef\tempurl%
\url{https://doi.org/10.1109/COMST.2023.3288942}
\showDOI{\tempurl}


\bibitem[Eltanbouly et~al\mbox{.}(2020)]%
        {eltanbouly2020}
\bibfield{author}{\bibinfo{person}{Sohaila Eltanbouly}, \bibinfo{person}{May
  Bashendy}, \bibinfo{person}{Noora AlNaimi}, \bibinfo{person}{Zina Chkirbene},
  {and} \bibinfo{person}{Aiman Erbad}.} \bibinfo{year}{2020}\natexlab{}.
\newblock \showarticletitle{Machine Learning Techniques for Network Anomaly
  Detection: A Survey}. In \bibinfo{booktitle}{\emph{2020 IEEE International
  Conference on Informatics, {IoT}, and Enabling Technologies (ICIoT)}}.
  \bibinfo{pages}{156--162}.
\newblock
\urldef\tempurl%
\url{https://doi.org/10.1109/ICIoT48696.2020.9089465}
\showDOI{\tempurl}


\bibitem[Engelen et~al\mbox{.}(2021)]%
        {cicflowmeter}
\bibfield{author}{\bibinfo{person}{Gints Engelen}, \bibinfo{person}{Vera
  Rimmer}, {and} \bibinfo{person}{Wouter Joosen}.}
  \bibinfo{year}{2021}\natexlab{}.
\newblock \showarticletitle{Troubleshooting an Intrusion Detection Dataset: the
  CICIDS2017 Case Study}. In \bibinfo{booktitle}{\emph{2021 IEEE Security and
  Privacy Workshops (SPW)}}. IEEE, \bibinfo{pages}{7--12}.
\newblock


\bibitem[Fleury et~al\mbox{.}(2015)]%
        {fleury_fit_2015}
\bibfield{author}{\bibinfo{person}{Eric Fleury}, \bibinfo{person}{Nathalie
  Mitton}, \bibinfo{person}{Thomas Noel}, {and} \bibinfo{person}{Cédric
  Adjih}.} \bibinfo{year}{2015}\natexlab{}.
\newblock \showarticletitle{{FIT} {IoT}-{LAB}: {The} {Largest} {IoT} {Open}
  {Experimental} {Testbed}}.
\newblock \bibinfo{journal}{\emph{ERCIM News}} \bibinfo{number}{101}
  (\bibinfo{date}{April} \bibinfo{year}{2015}), \bibinfo{pages}{4}.
\newblock
\urldef\tempurl%
\url{https://inria.hal.science/hal-01138038}
\showURL{%
\tempurl}


\bibitem[Ghazanfar et~al\mbox{.}(2020)]%
        {ghazanfar_iot-flock_2020}
\bibfield{author}{\bibinfo{person}{Syed Ghazanfar}, \bibinfo{person}{Faisal
  Hussain}, \bibinfo{person}{Atiq~Ur Rehman}, \bibinfo{person}{Ubaid~U.
  Fayyaz}, \bibinfo{person}{Farrukh Shahzad}, {and} \bibinfo{person}{Ghalib~A.
  Shah}.} \bibinfo{year}{2020}\natexlab{}.
\newblock \showarticletitle{{IoT}-{Flock}: {An} {Open}-source {Framework} for
  {IoT} {Traffic} {Generation}}.
\newblock \bibinfo{journal}{\emph{2020 International Conference on Emerging
  Trends in Smart Technologies (ICETST)}} (\bibinfo{date}{March}
  \bibinfo{year}{2020}), \bibinfo{pages}{1--6}.
\newblock
\showISBNx{9781728171135}
\urldef\tempurl%
\url{https://doi.org/10.1109/ICETST49965.2020.9080732}
\showDOI{\tempurl}
\newblock
\shownote{Conference Name: 2020 International Conference on Emerging Trends in
  Smart Technologies (ICETST)}.


\bibitem[Haseeb et~al\mbox{.}(2020)]%
        {haseeb_iot_killchain_2020}
\bibfield{author}{\bibinfo{person}{Junaid Haseeb}, \bibinfo{person}{Masood
  Mansoori}, {and} \bibinfo{person}{Ian Welch}.}
  \bibinfo{year}{2020}\natexlab{}.
\newblock \showarticletitle{A {Measurement} {Study} of {IoT}-{Based} {Attacks}
  {Using} {IoT} {Kill} {Chain}}.
\newblock \bibinfo{journal}{\emph{2020 IEEE 19th International Conference on
  Trust, Security and Privacy in Computing and Communications (TrustCom)}}
  (\bibinfo{date}{Dec.} \bibinfo{year}{2020}), \bibinfo{pages}{557--567}.
\newblock
\showISBNx{9781665403924}
\urldef\tempurl%
\url{https://doi.org/10.1109/TrustCom50675.2020.00080}
\showDOI{\tempurl}
\newblock
\shownote{Conference Name: 2020 IEEE 19th International Conference on Trust,
  Security and Privacy in Computing and Communications (TrustCom)}.


\bibitem[Helali(2020)]%
        {gns3_book}
\bibfield{author}{\bibinfo{person}{Saida Helali}.}
  \bibinfo{year}{2020}\natexlab{}.
\newblock \bibinfo{booktitle}{\emph{Simulating Network Architectures with
  GNS3}}.
\newblock \bibinfo{pages}{9--25}.
\newblock
\urldef\tempurl%
\url{https://doi.org/10.1002/9781119779964.ch2}
\showDOI{\tempurl}


\bibitem[Kambourakis et~al\mbox{.}(2017)]%
        {kambourakis_mirai_2017}
\bibfield{author}{\bibinfo{person}{Georgios Kambourakis},
  \bibinfo{person}{Constantinos Kolias}, {and} \bibinfo{person}{Angelos
  Stavrou}.} \bibinfo{year}{2017}\natexlab{}.
\newblock \showarticletitle{The {Mirai} botnet and the {IoT} {Zombie}
  {Armies}}. In \bibinfo{booktitle}{\emph{{MILCOM} 2017 - 2017 {IEEE}
  {Military} {Communications} {Conference} ({MILCOM})}}.
  \bibinfo{pages}{267--272}.
\newblock
\urldef\tempurl%
\url{https://doi.org/10.1109/MILCOM.2017.8170867}
\showDOI{\tempurl}
\newblock
\shownote{ISSN: 2155-7586}.


\bibitem[Kurimoto et~al\mbox{.}(2023)]%
        {kurimoto2023}
\bibfield{author}{\bibinfo{person}{Takashi Kurimoto}, \bibinfo{person}{Koji
  Sasayama}, \bibinfo{person}{Osamu Akashi}, {and} \bibinfo{person}{Shigeo
  Urushidani}.} \bibinfo{year}{2023}\natexlab{}.
\newblock \showarticletitle{SINET6: Nationwide 400GE-Based Academic Backbone
  Network in Japan}. \bibinfo{pages}{1--3}.
\newblock
\urldef\tempurl%
\url{https://doi.org/10.23919/OFC49934.2023.10117428}
\showDOI{\tempurl}


\bibitem[Moustafa et~al\mbox{.}(2023)]%
        {moustafa_xai_2023}
\bibfield{author}{\bibinfo{person}{Nour Moustafa}, \bibinfo{person}{Nickolaos
  Koroniotis}, \bibinfo{person}{Marwa Keshk}, \bibinfo{person}{Albert~Y.
  Zomaya}, {and} \bibinfo{person}{Zahir Tari}.}
  \bibinfo{year}{2023}\natexlab{}.
\newblock \showarticletitle{Explainable {Intrusion} {Detection} for {Cyber}
  {Defences} in the {Internet} of {Things}: {Opportunities} and {Solutions}}.
\newblock \bibinfo{journal}{\emph{IEEE Communications Surveys \& Tutorials}}
  \bibinfo{volume}{25}, \bibinfo{number}{3} (\bibinfo{year}{2023}),
  \bibinfo{pages}{1775--1807}.
\newblock
\showISSN{1553-877X}
\urldef\tempurl%
\url{https://doi.org/10.1109/COMST.2023.3280465}
\showDOI{\tempurl}
\newblock
\shownote{Conference Name: IEEE Communications Surveys \& Tutorials}.


\bibitem[of~Informatics(2021)]%
        {sinetstream2021}
\bibfield{author}{\bibinfo{person}{National~Institute of Informatics}.}
  \bibinfo{year}{2021}\natexlab{}.
\newblock \bibinfo{title}{SINETStream}.
\newblock
\newblock
\urldef\tempurl%
\url{https://www.sinetstream.net/index.en.html}
\showURL{%
\tempurl}
\newblock
\shownote{Japan. (Accessed on May 13th, 2024)}.


\bibitem[of~Informatics(2024)]%
        {sinet2024}
\bibfield{author}{\bibinfo{person}{National~Institute of Informatics}.}
  \bibinfo{year}{2024}\natexlab{}.
\newblock \bibinfo{title}{SINET6}.
\newblock
\newblock
\urldef\tempurl%
\url{https://www.sinet.ad.jp}
\showURL{%
\tempurl}
\newblock
\shownote{Japan. (Accessed on May 13th, 2024)}.


\bibitem[Osaki et~al\mbox{.}(2024)]%
        {ebpf_mirai}
\bibfield{author}{\bibinfo{person}{Atsuya Osaki}, \bibinfo{person}{Manuel
  Poisson}, \bibinfo{person}{Seiki Makino}, \bibinfo{person}{Shiiba Ryusei},
  \bibinfo{person}{Kensuke~Fukuda Fukuda}, \bibinfo{person}{Okoshi Tadashi},
  {and} \bibinfo{person}{Jin Nakazawa}.} \bibinfo{year}{2024}\natexlab{}.
\newblock \showarticletitle{Dynamic Fixed-point Values in eBPF: a Case for
  Fully In-kernel Anomaly Detection}.
\newblock \bibinfo{journal}{\emph{Proceedings of the 19th Asian Internet
  Engineering Conference}} (\bibinfo{date}{August} \bibinfo{year}{2024}),
  \bibinfo{pages}{9 pages}.
\newblock


\bibitem[Palmieri et~al\mbox{.}(2019)]%
        {palmieri_mqttsa_2019}
\bibfield{author}{\bibinfo{person}{Andrea Palmieri}, \bibinfo{person}{Paolo
  Prem}, \bibinfo{person}{Silvio Ranise}, \bibinfo{person}{Umberto Morelli},
  {and} \bibinfo{person}{Tahir Ahmad}.} \bibinfo{year}{2019}\natexlab{}.
\newblock \showarticletitle{{MQTTSA}: {A} {Tool} for {Automatically}
  {Assisting} the {Secure} {Deployments} of {MQTT} {Brokers}}. In
  \bibinfo{booktitle}{\emph{2019 {IEEE} {World} {Congress} on {Services}
  ({SERVICES})}}, Vol.~\bibinfo{volume}{2642-939X}. \bibinfo{pages}{47--53}.
\newblock
\urldef\tempurl%
\url{https://doi.org/10.1109/SERVICES.2019.00023}
\showDOI{\tempurl}
\newblock
\shownote{ISSN: 2642-939X}.


\bibitem[Patel et~al\mbox{.}(2019)]%
        {patel_simulators_2019}
\bibfield{author}{\bibinfo{person}{N~D Patel}, \bibinfo{person}{B~M Mehtre},
  {and} \bibinfo{person}{Rajeev Wankar}.} \bibinfo{year}{2019}\natexlab{}.
\newblock \showarticletitle{Simulators, {Emulators}, and {Test}-beds for
  {Internet} of {Things}: {A} {Comparison}}. In \bibinfo{booktitle}{\emph{2019
  {Third} {International} conference on {I}-{SMAC} ({IoT} in {Social},
  {Mobile}, {Analytics} and {Cloud}) ({I}-{SMAC})}}. \bibinfo{pages}{139--145}.
\newblock
\urldef\tempurl%
\url{https://doi.org/10.1109/I-SMAC47947.2019.9032519}
\showDOI{\tempurl}


\bibitem[{Rachit} et~al\mbox{.}(2021)]%
        {rachit_security_iot_trends_2021}
\bibfield{author}{\bibinfo{person}{{Rachit}}, \bibinfo{person}{Shobha Bhatt},
  {and} \bibinfo{person}{Prakash~Rao Ragiri}.} \bibinfo{year}{2021}\natexlab{}.
\newblock \showarticletitle{Security trends in {Internet} of {Things}: a
  survey}.
\newblock \bibinfo{journal}{\emph{SN Applied Sciences}} \bibinfo{volume}{3},
  \bibinfo{number}{1} (\bibinfo{date}{Jan.} \bibinfo{year}{2021}),
  \bibinfo{pages}{121}.
\newblock
\showISSN{2523-3971}
\urldef\tempurl%
\url{https://doi.org/10.1007/s42452-021-04156-9}
\showDOI{\tempurl}


\bibitem[Ring et~al\mbox{.}(2019)]%
        {ring2019}
\bibfield{author}{\bibinfo{person}{Markus Ring}, \bibinfo{person}{Sarah
  Wunderlich}, \bibinfo{person}{Deniz Scheuring}, \bibinfo{person}{Dieter
  Landes}, {and} \bibinfo{person}{Andreas Hotho}.}
  \bibinfo{year}{2019}\natexlab{}.
\newblock \showarticletitle{A survey of network-based intrusion detection data
  sets}.
\newblock \bibinfo{journal}{\emph{Computers \& Security}}  \bibinfo{volume}{86}
  (\bibinfo{year}{2019}), \bibinfo{pages}{147--167}.
\newblock
\showISSN{0167-4048}
\urldef\tempurl%
\url{https://doi.org/10.1016/j.cose.2019.06.005}
\showDOI{\tempurl}


\bibitem[Saez-de Camara et~al\mbox{.}(2023)]%
        {saez-de-camara_gotham_2023}
\bibfield{author}{\bibinfo{person}{Xabier Saez-de Camara},
  \bibinfo{person}{Jose~Luis Flores}, \bibinfo{person}{Cristóbal Arellano},
  \bibinfo{person}{Aitor Urbieta}, {and} \bibinfo{person}{Urko Zurutuza}.}
  \bibinfo{year}{2023}\natexlab{}.
\newblock \showarticletitle{Gotham {Testbed}: {A} {Reproducible} {IoT}
  {Testbed} for {Security} {Experiments} and {Dataset} {Generation}}.
\newblock \bibinfo{journal}{\emph{IEEE Transactions on Dependable and Secure
  Computing}}  \bibinfo{volume}{PP} (\bibinfo{date}{Jan.}
  \bibinfo{year}{2023}), \bibinfo{pages}{1--18}.
\newblock
\urldef\tempurl%
\url{https://doi.org/10.1109/TDSC.2023.3247166}
\showDOI{\tempurl}


\bibitem[Shirsath(2023)]%
        {shirsath_caida_2023}
\bibfield{author}{\bibinfo{person}{Vaishali Shirsath}.}
  \bibinfo{year}{2023}\natexlab{}.
\newblock \bibinfo{title}{{CAIDA} {UCSD} {DDoS} 2007 {Attack} {Dataset}}.
\newblock
\newblock
\urldef\tempurl%
\url{https://ieee-dataport.org/documents/caida-ucsd-ddos-2007-attack-dataset}
\showURL{%
\tempurl}
\newblock
\shownote{(Accessed on May 13th, 2024)}.


\bibitem[Singh et~al\mbox{.}(2022)]%
        {singh_survey_simulation_2022}
\bibfield{author}{\bibinfo{person}{Abhishek Singh}, \bibinfo{person}{Himanshu
  Nandanwar}, {and} \bibinfo{person}{Anamika Chauhan}.}
  \bibinfo{year}{2022}\natexlab{}.
\newblock \showarticletitle{Simulation {Tools} and {Testbeds} for {Internet} of
  {Things}({IoT}): “{Comparative} {Insight}”}. In
  \bibinfo{booktitle}{\emph{2022 {Second} {International} {Conference} on
  {Computer} {Science}, {Engineering} and {Applications} ({ICCSEA})}}.
  \bibinfo{pages}{1--7}.
\newblock
\urldef\tempurl%
\url{https://doi.org/10.1109/ICCSEA54677.2022.9936302}
\showDOI{\tempurl}


\bibitem[Takefusa et~al\mbox{.}(2021)]%
        {takefusa_sinetstream_2021}
\bibfield{author}{\bibinfo{person}{Atsuko Takefusa}, \bibinfo{person}{Jingtao
  Sun}, \bibinfo{person}{Ikki Fujiwara}, \bibinfo{person}{Hiroshi Yoshida},
  \bibinfo{person}{Kento Aida}, {and} \bibinfo{person}{Calton Pu}.}
  \bibinfo{year}{2021}\natexlab{}.
\newblock \showarticletitle{{SINETStream}: {Enabling} {Research} {IoT}
  {Applications} with {Portability}, {Security} and {Performance}
  {Requirements}}. In \bibinfo{booktitle}{\emph{2021 {IEEE} 45th {Annual}
  {Computers}, {Software}, and {Applications} {Conference} ({COMPSAC})}}.
  \bibinfo{pages}{482--492}.
\newblock
\urldef\tempurl%
\url{https://doi.org/10.1109/COMPSAC51774.2021.00073}
\showDOI{\tempurl}
\newblock
\shownote{ISSN: 0730-3157}.


\bibitem[Terroso-Saenz et~al\mbox{.}(2019)]%
        {terroso_saenz_hybrid_testbed_2019}
\bibfield{author}{\bibinfo{person}{Fernando Terroso-Saenz},
  \bibinfo{person}{Aurora González-Vidal}, \bibinfo{person}{Alfonso~P.
  Ramallo-González}, {and} \bibinfo{person}{Antonio~F. Skarmeta}.}
  \bibinfo{year}{2019}\natexlab{}.
\newblock \showarticletitle{An open {IoT} platform for the management and
  analysis of energy data}.
\newblock \bibinfo{journal}{\emph{Future Generation Computer Systems}}
  \bibinfo{volume}{92} (\bibinfo{date}{March} \bibinfo{year}{2019}),
  \bibinfo{pages}{1066--1079}.
\newblock
\showISSN{0167-739X}
\urldef\tempurl%
\url{https://doi.org/10.1016/j.future.2017.08.046}
\showDOI{\tempurl}


\bibitem[Vaccari et~al\mbox{.}(2020a)]%
        {vaccari_slowite_2020}
\bibfield{author}{\bibinfo{person}{Ivan Vaccari}, \bibinfo{person}{Maurizio
  Aiello}, {and} \bibinfo{person}{Enrico Cambiaso}.}
  \bibinfo{year}{2020}\natexlab{a}.
\newblock \showarticletitle{{SlowITe}, a {Novel} {Denial} of {Service} {Attack}
  {Affecting} {MQTT}}.
\newblock \bibinfo{journal}{\emph{Sensors}}  \bibinfo{volume}{20}
  (\bibinfo{date}{May} \bibinfo{year}{2020}), \bibinfo{pages}{2932}.
\newblock
\urldef\tempurl%
\url{https://doi.org/10.3390/s20102932}
\showDOI{\tempurl}


\bibitem[Vaccari et~al\mbox{.}(2020b)]%
        {vaccari_mqttset_2020}
\bibfield{author}{\bibinfo{person}{Ivan Vaccari}, \bibinfo{person}{Giovanni
  Chiola}, \bibinfo{person}{Maurizio Aiello}, \bibinfo{person}{Maurizio
  Mongelli}, {and} \bibinfo{person}{Enrico Cambiaso}.}
  \bibinfo{year}{2020}\natexlab{b}.
\newblock \showarticletitle{{MQTTset}, a {New} {Dataset} for {Machine}
  {Learning} {Techniques} on {MQTT}}.
\newblock \bibinfo{journal}{\emph{Sensors}} \bibinfo{volume}{20},
  \bibinfo{number}{22} (\bibinfo{date}{Jan.} \bibinfo{year}{2020}),
  \bibinfo{pages}{6578}.
\newblock
\showISSN{1424-8220}
\urldef\tempurl%
\url{https://doi.org/10.3390/s20226578}
\showDOI{\tempurl}
\newblock
\shownote{Number: 22 Publisher: Multidisciplinary Digital Publishing
  Institute}.


\bibitem[Zhu et~al\mbox{.}(2022)]%
        {zhu_survey_2022}
\bibfield{author}{\bibinfo{person}{Shicheng Zhu}, \bibinfo{person}{Shunkun
  Yang}, \bibinfo{person}{Xiaodong Gou}, \bibinfo{person}{Yang Xu},
  \bibinfo{person}{Tao Zhang}, {and} \bibinfo{person}{Yueliang Wan}.}
  \bibinfo{year}{2022}\natexlab{}.
\newblock \showarticletitle{Survey of {Testing} {Methods} and {Testbed}
  {Development} {Concerning} {Internet} of {Things}}.
\newblock \bibinfo{journal}{\emph{Wireless Personal Communications}}
  \bibinfo{volume}{123}, \bibinfo{number}{1} (\bibinfo{date}{March}
  \bibinfo{year}{2022}), \bibinfo{pages}{165--194}.
\newblock
\showISSN{1572-834X}
\urldef\tempurl%
\url{https://doi.org/10.1007/s11277-021-09124-5}
\showDOI{\tempurl}


\end{thebibliography}
